\documentclass[11pt,twoside]{article}

\usepackage{asp2006}
\usepackage{graphicx}
\usepackage{epsf}

\begin{document}
\title{Does the Selection of a Quiet Region Influence the Local Helioseismic
Inferences?}   
\author{Sushanta C. Tripathy,\altaffilmark{1} H. M. Antia,\altaffilmark{2}
Kiran Jain,\altaffilmark{1} and Frank Hill\altaffilmark{1}}   
\altaffiltext{1}{National Solar Observatory, 950 N. Cherry Ave.,  Tucson, AZ
85719, USA}  
\altaffiltext{2}{Tata Institute of Fundamental Research, Homi Bhabha Road,
Mumbai 400 005, India}

\begin{abstract}
We apply the ring-diagram technique to high resolution Dopplergrams in order to
estimate the variation in oscillation mode parameters between active and quiet
regions.  We demonstrate that the difference in
mode parameters between two quiet regions can be as large as those between a
pair of active and quiet region. This leads us to conclude that the results
derived on the basis of a single quiet region could be biased. 
\end{abstract}

\section{Introduction}  
In local helioseismic studies the oscillation mode parameters of an active
region are often compared with those of a quiet region to estimate the
influence of
the magnetic field or differences in structure. There are various ways in
which a quiet region can be selected: (i) a common quiet region for all the
events analyzed (ii) a quiet region at the same heliographic longitude and
latitude and within the same Carrington rotation, and (iii) an ensemble average
of quiet regions. The first choice minimizes the differences that may arise
from selection of different quiet regions but neglects the effect of temporal
and spatial 
variations. The second choice has been used in several studies (e.g.
\citeauthor{antia01} \citeyear{antia01})
assuming that the differences in mode parameters between two quiet regions are
small compared to a pair of  active and quiet region. Here we investigate how
the selection procedure affects the variations in inferred 
mode parameters.  

\section{Data and Methodology}
To analyze the properties of the modes in quiet and active regions, we use the
technique of ring diagrams \citep{hill88} where we analyze a time
series of  Dopplergrams.  Each region of about 15\deg\ $\times$ 15\deg\ in
heliographic longitude and latitude is tracked with the mean surface rotation
velocity for 1664 minutes giving a frequency resolution of about 10~$\mu$Hz.
The three-dimensional power spectrum of the time series is fitted to obtain 
the mode parameters and flow velocities. More details of the procedure 
can be found in \citet{ring}.

For this investigation, we have identified a total of
42 active regions during the period of September 2001 to December 2003 and use
the high-resolution Global Oscillation Network Group (GONG+) Dopplergrams to
obtain the mode parameters.  We also use Michelson Doppler Imager (MDI)
magnetograms to calculate the Magnetic
Activity Index (MAI) which is a measure of the strength of the strong field
component of the associated magnetic field \citep[see][for a detailed
description]{antia01}. The MAI is calculated from the  96-minute magnetograms
by integrating the unsigned, strong-field values
within the selected regions and over the same time intervals used to track and
calculate the power spectra. 

\begin{figure}[!t]
\begin{center}
\includegraphics[width=0.27\linewidth,angle=90]{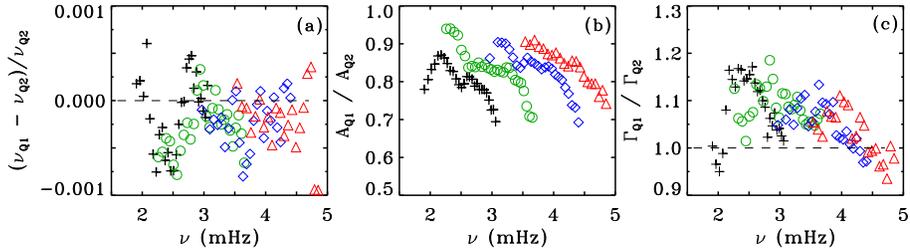}
\caption{Changes in (a) mode frequency (b) amplitude and (c) line width of
oscillation modes between two quiet regions $Q1$ and $Q2$ within CR 1998. Both
of the regions are located at S22E25 corresponding to AR NOAA 10224 on 2003
January 7 and have MAI values of 2.06~G and 2.08~G, respectively. The $f$ modes
are presented as pluses and $p$ modes are presented as circles, diamonds, and
triangles for $p_1$ through $p_3$, respectively.   \label{fig1}}
\end{center}
\end{figure}
\section{Results and Discussions}
To illustrate the variations in mode parameters between quiet regions, we
select two quiet regions at the same
latitude and within Carrington rotation 1998 corresponding to  active
region (AR) NOAA 10224 located at S22E25 on 2003 January 7 (MAI = 16.22~G).
These regions (hereafter
$Q1$ and $Q2$) are located at Carrington longitude  62.5\deg\ and  
7.5\deg\ and have  MAI values of 2.06 G and 2.08 G, respectively.    
Figure~\ref{fig1} shows  the variations of different mode parameters. It
is evident that the relative frequency differences  (Figure~1a)
are small and  consistent with differences expected for
low MAI regions. Figure 1b and 1c shows the amplitudes and line widths    
 as a ratio between $Q1$ and $Q2$, respectively.  A variation 
of about 20--30\% is observed in both of the parameters which is not expected
to be seen 
between two quiet regions having similar MAIs. This illustrates the intrinsic
uncertainties associated with quiet regions and reveals the importance of
selecting appropriate quiet regions for comparison studies.

\begin{figure}[!t]
\begin{center}
\includegraphics[width=0.60\linewidth]{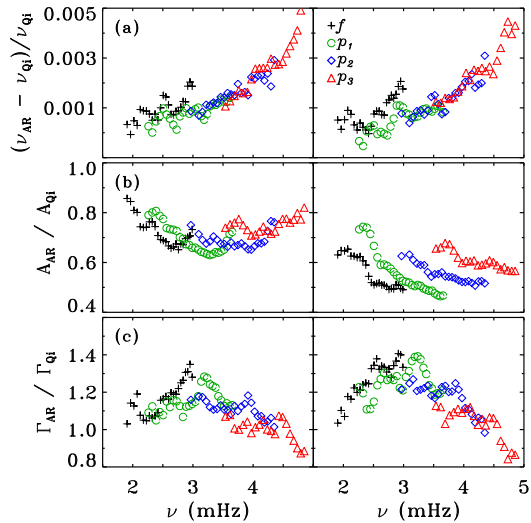}
\caption{(a) Relative frequency differences, (b) ratio of amplitudes, and (c)
ratio of line widths of oscillation modes between AR 10244 and quiet regions
$Q1$ (left panels) and $Q2$ (right panels). \label{fig2}}
\end{center}
\end{figure}

To understand how the inferred mode properties of the AR can be  affected by
the choice of
the quiet regions, we plot the same three parameters between AR
 with respect to $Q1$ and $Q2$ in Figure~2. Figure~\ref{fig2}a depicts the
relative frequency differences and  we do not see  
significant changes between the two panels. Figure~\ref{fig2}b
shows the  amplitude as a ratio
between the AR and quiet regions and in general it is found that the power in
the AR is suppressed compared to the quiet Sun. However, on a closer
examination some differences are noticed. 
When the comparison is carried out with respect to $Q1$ (left panel) the
maximum
power suppression occurs in the frequency range
of about 3--3.5~mHz while the comparison 
with $Q2$ (right panel) shows that the suppression decreases monotonically 
with frequency.  

Figure~\ref{fig2}c shows the ratio of line widths between the AR and quiet
regions and   
indicates significant changes in widths both as a function of radial 
order and frequency.  In general, we find that  the line widths of AR
are higher compared to the quiet regions. 
We also notice a few other subtle differences between the two panels, for
example 
the right panel shows a 5--10\% larger width compared to the left panel.  
This provides evidence that 
the changes in mode parameters of an active region can also be affected by the
choice of the quiet regions.

\begin{figure}[!t]
\begin{center}
\plottwo{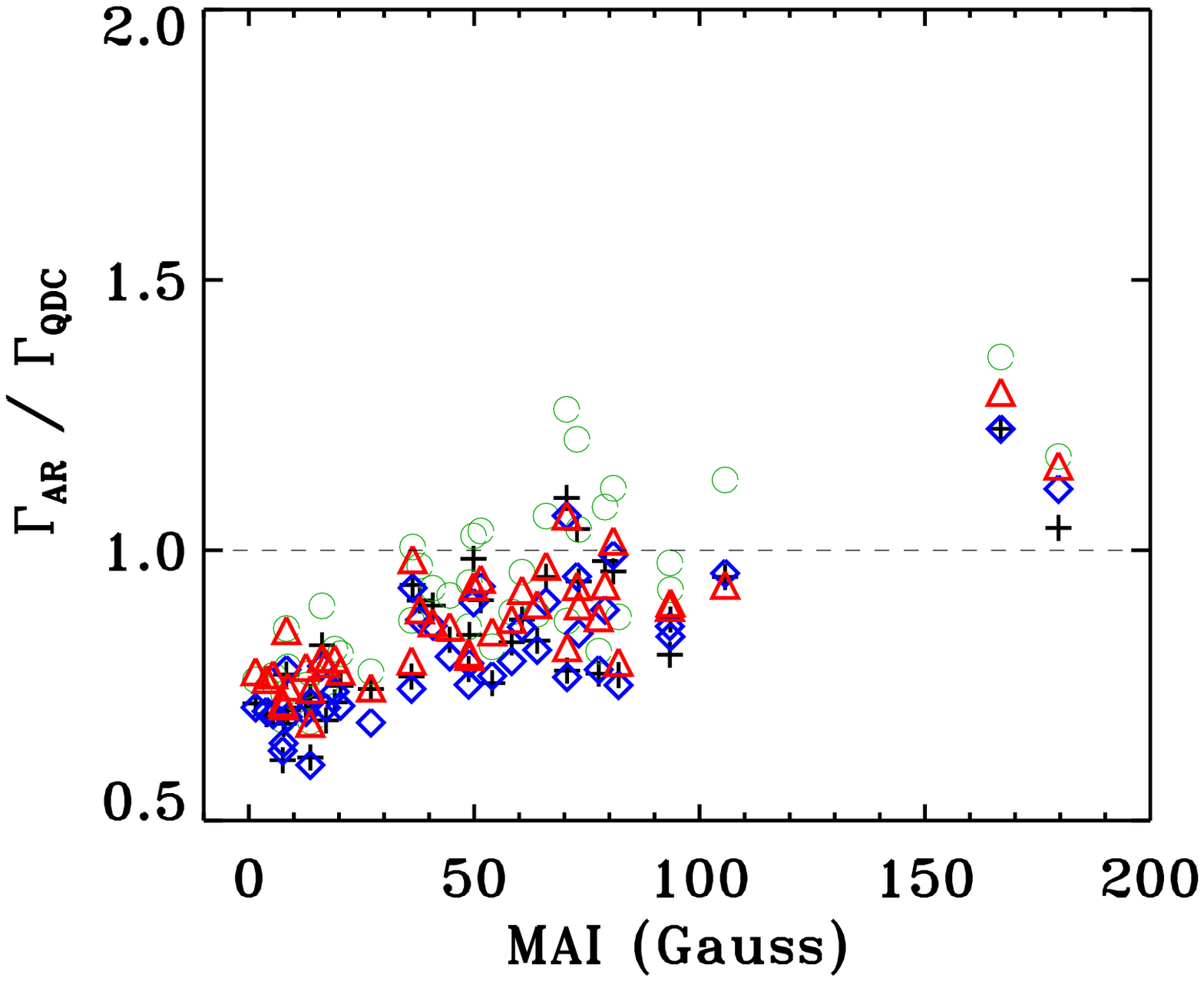}{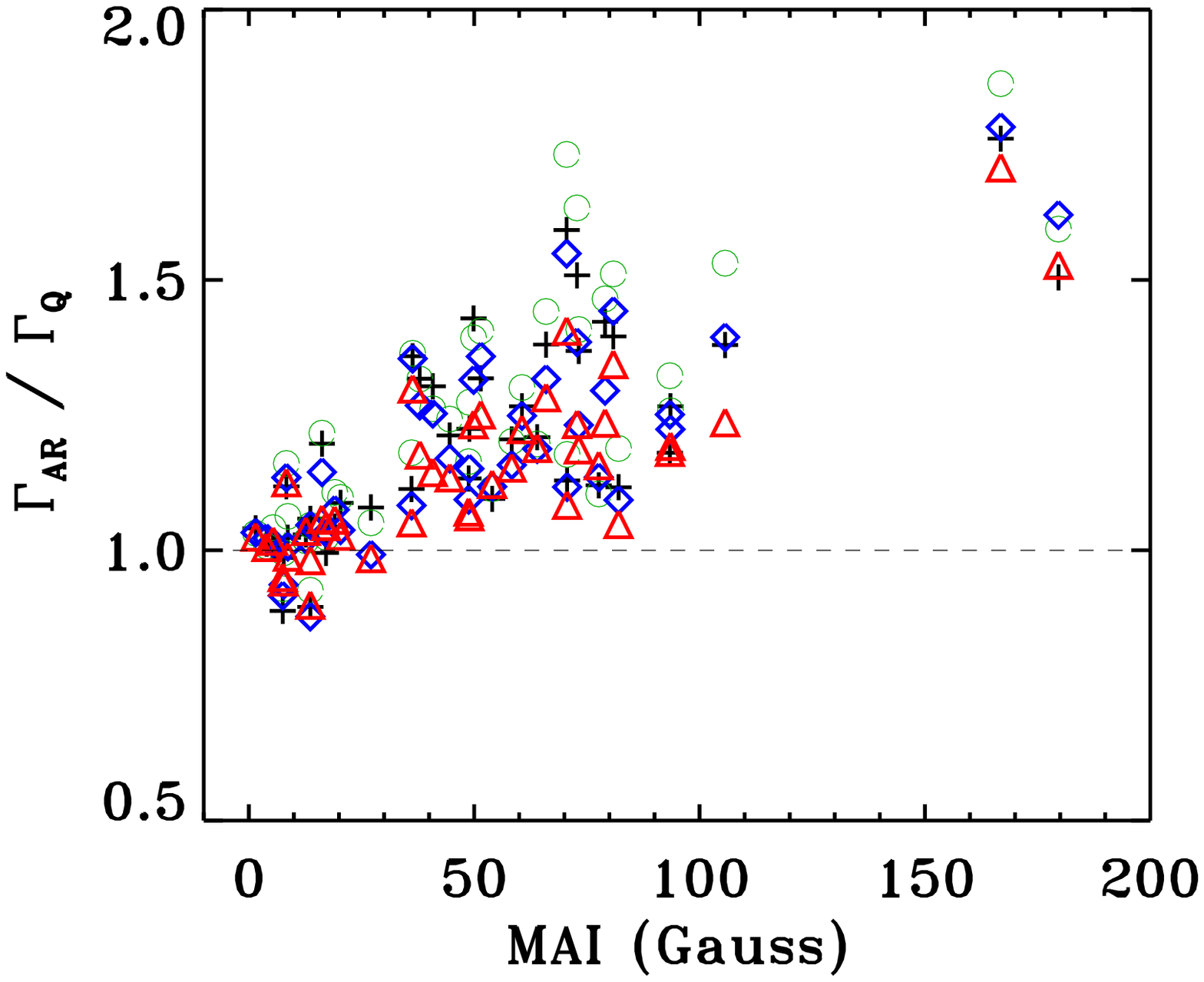}
\caption{Frequency-averaged line widths as a function of the MAI of active
regions. The left panel denotes the ratio with respect to QDC 
while the right panel shows the ratio with respect to
the ensemble average ($Q$).  The symbols have the same meaning as in Figure 2.
\label{fig3}}
\end{center}
\end{figure}

In order to calculate an ensemble average of quiet regions which can then be
used to compare the mode properties of all 42 active regions, we selected
one quiet region at the disk center per Carrington rotation covered by the
active regions. The MAI of these quiet regions varied between 0.387~G to
7.738~G. However, to minimize 
the effect due to  intrinsic variations between different quiet regions, 
we averaged together the mode properties of those quiet regions that have a
MAI less than 3~G. Only 22 quiet regions satisfied this criterion and the
average mode parameters of these quiet regions are defined as the parameters
of the control quiet region, ($Q$) with an average MAI of 0.93 G.  As a second
choice, we
select a quiet region at the disk center (QDC) on 2002 May 19 since this period 
roughly corresponds to the middle of the data sets; the region has a MAI of
2.78~G.  In
Figure~\ref{fig3}, we compare the frequency-averaged line widths as a function
of the MAI of ARs for these two different selection of quiet regions. The
$f$ modes have been averaged over  the frequency range of 2500--2750
$\mu$Hz, $p_1$ and $p_2$ over the range of 3000--3500 $\mu$Hz and
$p_3$ over  the range of 3250--3750 $\mu$Hz. The left panel denotes
the ratio with respect to the quiet region, QDC while the right
panel shows the ratio with respect to the ensemble average, $Q$. The 
behavior in line widths  between the two panels are distinctly different.
While the left panel shows smaller widths, the right panel shows higher widths 
except for widths associated with low MAIs.  Therefore it is clear that the
inferred properties of oscillation modes, specifically, the line widths, 
depend 
on the choice of the quiet region and can strongly influence the conclusions.

In summary, we find that the mode amplitudes and line widths between two quiet
regions located within one Carrington rotation can show a large variation.
Thus,  we conclude that the choice of a single
quiet region can mislead the interpretation and we propose that future studies
involving comparison of mode properties between a pair of active and quiet
region use an ensemble average of quiet regions as an optimum choice. 

\acknowledgements 
We thank John Leibacher for useful comments. This research 
was supported in part by NASA grant NNG 05HL41I and NNG 08EI54I.
This work utilizes data obtained by the Global Oscillation Network
Group program, managed by the National Solar Observatory, which
is operated by AURA, Inc. under a cooperative agreement with the
National Science Foundation. The data were acquired by instruments
operated by the Big Bear Solar Observatory, High Altitude Observatory,
Learmonth Solar Observatory, Udaipur Solar Observatory, Instituto de
Astrof\'{\i}sica de Canarias, and Cerro Tololo Interamerican
Observatory.  SOHO is a  project of international cooperation between ESA and NASA.

\end{document}